\documentclass[aps,showpacs,preprintnumbers,showkeys, amsmath, amssymb]{revtex4}
\usepackage{float}
\usepackage{graphics,epsfig}
\usepackage{graphicx}
\usepackage{dcolumn}
\usepackage{bm}

\begin{document}
\baselineskip=0.65 cm

\title{ Dynamics and quantum entanglement of two-level atoms in de Sitter spacetime}
\author{Zehua Tian and Jiliang {Jing}\footnote{Corresponding author, electronic address:
jljing@hunnu.edu.cn}}

\affiliation{Department of Physics, and Key Laboratory of Low Dimensional Quantum Structures and
Quantum Control of Ministry of Education, Hunan Normal University,
Changsha, Hunan 410081, P. R. China}

\begin{abstract}
\baselineskip=0.55 cm
\begin{center}
{\bf Abstract}
\end{center}

In the framework of open quantum systems, we study the internal dynamics of both freely falling and static two-level atoms interacting with quantized conformally coupled massless scalar field in de Sitter spacetime. We find that the atomic transition rates depend on both the nature of de Sitter spacetime and the motion of atoms, interestingly the steady states for both cases are always driven to being purely thermal, regardless of the atomic initial states. This thermalization phenomenon is structurally similar to what happens to an elementary quantum system immersed in a thermal field, and thus reveals the thermal nature of de Sitter spacetime. Besides, we find that the thermal baths will drive the entanglement shared by the freely falling atom (the static atom) and its auxiliary partner, a same two-level atom which is isolated from external fields, to being sudden death, and the proper time for the entanglement to be extinguished is computed. We also analyze that such thermalization and disentanglement phenomena, in principle, could be understood from the perspective of table-top simulation experiment.

\end{abstract}



\maketitle
\newpage

\section{Introduction}

One of the amazing effects predicted by the relativistic quantum field theory is the Unruh effect \cite{Unruh,Fulling,Birrell}. It is represented as that a uniformly accelerated observer can view an extraordinary phenomenon, thermal particles, in Minkowski vacuum, which is completely contrary to the inertial status. Usually, the Unruh effect, in theory, can be read from the excitation rate, i.e, the probability per unit time of a spontaneous transition from the ground state to one of its excited states, of a uniformly accelerated DeWitt detector \cite{Birrell,Crispino} due to the appearance of Planck factor in it. This factor is considered as the concrete embodiment of thermal effect, because it indicates that the equilibrium between the accelerated detector and the external field is the same as that which would have been achieved had the detector remained unaccelerated, but immersed in a bath of thermal radiation at a temperature associated with its proper acceleration, which is the exact Unruh temperature \cite{Birrell}. Similar analysis also can be done to discuss Hawking effect \cite{Hawking}, one of the famous predicted phenomena in curved spacetime.

Besides the detector method, a wonderful alternative scheme to study the Unruh effect and Hawking effect is the open quantum system method. This method involves many important areas in physics, such as quantum mechanics, quantum field theory and relativistic theory. Therefore, it has attracted much attention recently \cite{Benatti,Yu1,Tian,Hu}. A system, such as a uniformly accelerated two-level atom, which couples with the external field in Minkowski vacuum, is considered as an open quantum system. Through studying its evolution, we will find that the density matrix corresponding to the quantum system is eventually driven to a purely thermal equilibrium state, and exhibits a nonvanishing probability of spontaneous excitation. This phenomenon is usually referred to as the Unruh effect. It is needed to note that this approach has been extended to understanding Hawking effect \cite{Yu1} in curved spacetime by assuming a two-level atom in the interaction with vacuum fluctuations. Besides, this method has also been used to analyze the geometric phase of  a two-level atom to detect the Unruh temperature \cite{Hu} and reveal the nature of de Sitter spacetime \cite{Tian}. In this regard, let us note that the theory of open quantum system has been fruitfully applied to studying relativistic effects \cite{Benatti,Yu1,Tian,Hu,Doukas}.

A two-level atom, which interacts with quantized conformally coupled massless scalar field in de Sitter-invariant vacuum, can be thought of as an open quantum system, and the massless scalar field it interacts with is equivalent to external the environment. One may expect that the evolution of the atom, e.g, transition  rates between energy levels and the thermalization process, will be influenced by the spacetime curvature which backscatters the fluctuating scalar field the atom is coupled to. Indeed, further study shows that both the freely falling atom and static atom in weak interaction with a massless scalar field in de Sitter spacetime feel a thermal bath and will be subjected to dissipation \cite{Tian}. Moreover, by introducing an auxiliary system (the same two-level atom, we call it the auxiliary partner), which is initially entangled with our freely falling atom or static atom and isolated from external field, we can discuss how the nature of de Sitter spacetime affects the dynamic evolution of bipartite atomic entanglement, a very important quantum resource, which plays a key role in the quantum information tasks such as quantum teleportation \cite{Bennett,Horodecki} and computation \cite{Horodecki}. It is needed to note that this model is in structural similarity to a bipartite quantum system in quantum information theory, with one subsystem in interaction with external environment, and the other isolated from that. In this regard, let us note that this model has been used to discuss the loss of spin entanglement for accelerated electrons in electric magnetic fields \cite{Doukas}, and the entanglement of two qubits in a relativistic orbit \cite{Carson}.

It is well known that entanglement is observer dependent \cite{Alsing}, incorporating the concepts of quantum information into relativistic settings can produce new and surprising effects \cite{Mann}. For example, from the perspective of the accelerated observer the entanglement shared no matter by the Bosonic fields or by the Dirac fields decreases with the increase of acceleration. Furthermore, because of different statistics, the entanglement for the Bosonic fields will disappear in the limit of infinity acceleration, while it is not so for the Dirac fields \cite{Fuentes,Tessier}. Recently, the generation of quantum entanglement for both the Bosonic fields and Dirac fields has been investigated due to the expansion of the universe \cite{Ball,Moradi}, and Mart\'{\i}n-Mart\'{\i}nez et. al. gave a review on the cosmological quantum entanglement \cite{Martin}. It is needed to note that the entanglement studied in previous articles \cite{Fuentes,Tessier,Ball,Moradi} is the entanglement of the free field modes, and these free field modes are spatially delocalized, which cannot be measured and processed. So it is remains interesting to see what happens to the spatially localized entangled quantum system when it is subjected to the spacetime effect, such as Gibbons-Hawking effect \cite{Gibbons}, due to being placed in a curved spacetime. To aim at that, we, in this paper, study the dynamics and entanglement for two-level atoms in de Sitter spacetime. The reason for our special attention to de Sitter spacetime stems from the fact that de Sitter space is the unique maximally symmetric curved spacetime. It enjoys the same degree of symmetry as Minkowski space (ten Killing vectors). Besides, the current observations, together with the theory of inflation, suggest that our universe may have approached de Sitter geometries in the far past and may approach de Sitter geometries in the far future. We respectively study the dynamics of both the freely falling atom and the static atom in de Sitter spacetime, and analyze the entanglement shared by the freely falling atom (static atom) with its auxiliary partner. For both cases, we find that the dissipation effect of thermal bath that the freely falling atom (static atom) feels will eventually drive the maximally entangled state to a separable state, to be more precise, a
classical state.

Our paper is constructed as follows: after briefly reviewing quantum evolution of two-level atoms and simply introducing the concurrence of bipartite quantum system in Section \ref{section 1}, we calculate and discuss the concurrence of the freely falling atom with its auxiliary partner in de Sitter spacetime in Section \ref{section 2} and that of the static atom with its auxiliary partner in Section \ref{section 3}. In Section \ref{section 4} and Section \ref{section 5} we give our discussions and conclusions, respectively.

\section{Dynamic evolution of two-level atom and introduction of concurrence} \label{section 1}

In this section, we will study the dynamic evolution of two-level atom, and give a brief introduction to the concurrence, a generally used measurement to quantify entanglement.

\subsection{Dynamic evolution of single two-level atom system}

Let us begin with the Hamiltonian of the total system, atom plus external field, which is given by
\begin{eqnarray}\label{Hamiltonian}
H=H_s+H_\phi+H_I,
\end{eqnarray}
where $H_s$ and $H_\phi$ are the Hamiltonian of atom and scalar field, and $H_I$ represents their interaction. For simplicity, we take a two-level atom with Hamiltonian $H_s=\frac{1}{2}\omega_0\sigma_z$, where $\omega_0$ is the energy level spacing of the atom, and $\sigma_z$ is the Pauli matrix. We assume that the Hamiltonian describing the interaction between atom and scalar field is $H_I=\mu(\sigma_++\sigma_-)\phi(x(\tau))$, in which $\mu$ is the coupling constant, $\sigma_+$ ($\sigma_-$) is the atomic rasing (lowering) operator, and $\phi(x)$ corresponds to the scalar field operator in de Sitter spacetime.

Initially the total density operator of the system plus field is assumed to be $\rho_{tot}=\rho(0)\otimes|0\rangle\langle0|$, in which $\rho(0)$ is the reduced density matrix of the atom, and $|0\rangle$ is the vacuum of the field. For the total system, its equation of motion is given by
\begin{eqnarray}\label{motion equation}
\frac{\partial\rho_{tot}(\tau)}{\partial\tau}=-i[H,\rho_{tot}(\tau)],
\end{eqnarray}
where $\tau$ is the proper time of the atom. In the limit of weak coupling, the evolution of the reduced density matrix $\rho(\tau)$, after simplification, can be written in the Lindblad form \cite{Lindblad,Benatti}
\begin{eqnarray}\label{Lindblad equation}
\nonumber
\frac{\partial\rho(\tau)}{\partial\tau}&=&-i[H_{eff},\rho(\tau)]+\cal L[\rho(\tau)]
\\
&=&-i[H_{eff},\rho(\tau)]+\sum^3_{j=1}[2L_j\rho L^\dagger_j-L^\dagger_jL_j\rho-\rho L^\dagger_jL_j],
\end{eqnarray}
where $H_{eff}$ and $L_j$ are given by
\begin{eqnarray}\label{L values}
\nonumber
H_{eff}=\frac{1}{2}\Omega\sigma_z=\frac{1}{2}\{\omega_0+\mu^2\mathrm{Im}(\Gamma_++\Gamma_-)\}\sigma_z,
\\
L_1=\sqrt{\frac{\gamma_-}{2}}\sigma_-,~~L_2=\sqrt{\frac{\gamma_+}{2}}\sigma_+,
~~L_3=\sqrt{\frac{\gamma_z}{2}}\sigma_z,
\end{eqnarray}
with
\begin{eqnarray}
\nonumber
\gamma_\pm&=&2\mu^2\mathrm{Re}\Gamma_\pm=\mu^2\int^{+\infty}_{-\infty}e^{\mp i\omega_0s}G^+(s-i\epsilon)ds,\nonumber\\
\gamma_z&=&0.
\end{eqnarray}
$G^+(x-x')=\langle0|\phi(x)\phi(x')|0\rangle$ is the field correlation function and $s=\tau-\tau'$ here.

For a single two-level atom with initial state $|\psi(0)\rangle=\cos\frac{\theta}{2}|1\rangle+\sin\frac{\theta}{2}|0\rangle$, its time-dependent reduced density matrix, according to Eq. (\ref{Lindblad equation}), is given by
\begin{eqnarray}\label{time-dependent state}
&&\rho(\tau)=\nonumber \\ && \frac{1}{2}
\left(
\begin{array}{cc}
1+e^{-(\gamma_++\gamma_-)\tau}\cos\theta+\frac{\gamma_+-\gamma_-} {\gamma_++\gamma_-}(1-e^{-(\gamma_++\gamma_-)\tau})& e^{-\frac{1}{2}(\gamma_++\gamma_-)\tau-i\Omega\tau}\sin\theta \\
e^{-\frac{1}{2}(\gamma_++\gamma_-)\tau+i\Omega\tau}\sin\theta & 1-e^{-(\gamma_++\gamma_-)\tau}\cos\theta-\frac{\gamma_+-\gamma_-} {\gamma_++\gamma_-}(1-e^{-(\gamma_++\gamma_-)\tau}) \\
\end{array}
\right).
\end{eqnarray}
Eq. (\ref{time-dependent state}) shows the effects of decoherence and dissipation on the atom, $\frac{\gamma_++\gamma_-}{2}$ describes the time scale for the off-diagonal elements of the density-matrix (``coherence") decay and $\gamma_++\gamma_-$ is the time scale for atomic transition. For $\tau\gg1/(\gamma_++\gamma_-)$, i.e., sufficiently long period of time for atomic evolution, the atom will be driven to a steady state,
\begin{eqnarray}\label{steady state}
\rho(\tau)=\frac{1}{\gamma_++\gamma_-}
\left(
\begin{array}{cc}
\gamma_+ & 0 \\
0 & \gamma_- \\
\end{array}
\right),
\end{eqnarray}
where $\gamma_++\gamma_-$ is the total transition rate. Then the ratio of the transition rates is $\frac{\gamma_+}{\gamma_-}$, which defines an equilibrium ratio of populations of the upper and lower
states. This equilibrium distribution over the levels could always tell us some information of the atomic steady state, such as thermal or non-thermal.

%%%%%%%%%%%%%%%%%%%%%%%%%%%%%%%%%%%%%%%%%%%%%%%%%%%%%%%%%%%%%%%%%%

\subsection{Brief introduction to concurrence}

Concurrence, which is always used to measure entanglement, is defined as \cite{Wootters,Wootters1}
\begin{eqnarray}\label{definition of concurrence}
C(\rho)=\max\{\lambda_1-\lambda_2-\lambda_3-\lambda_4,0\},~~~\lambda_i\geq\lambda_{i+1}\geq0.
\end{eqnarray}
where $\{\lambda_1,\lambda_2,\lambda_3,\lambda_4\}$ are square roots of the eigenvalues of the matrix
$\rho\widetilde{\rho}_s$ with $\widetilde{\rho}_s=(\sigma_y\otimes\sigma_y)\rho^\ast(\sigma_y\otimes\sigma_y)$.

To discuss the dynamic evolution of entanglement, we introduce an auxiliary system (a same two-level atom) which is isolated from external environment\cite{Doukas}. After doing like this, $\rho$ spans a sixteen dimensional vector space and the direct product of Pauli matrices including the identity, $\{\sigma_i\otimes\sigma_j|i,j\in0,...,3\}$, forms sixteen linearly independent vectors, which can expand any general density matrix for the bipartite two-level atom system introduced above. For convenience, we write the density matrix of two atoms $\rho(\tau)$ in terms of the Pauli matrices
\begin{eqnarray}\label{density matrices}
\rho(\tau)=\sum^3_{i=0}\sum^3_{j=0}\rho_{ij}(\tau)\sigma_i\otimes\sigma_j.
\end{eqnarray}
By substituting Eq. (\ref{density matrices}) into Eq. (\ref{Lindblad equation}), the time dependent state parameters,
after a series of calculations, are found to be
\begin{eqnarray}\label{solutions}
 \nonumber
\rho_{0j}(\tau)&=&\rho_{0j}(0),
\\  \nonumber
\rho_{1j}(\tau)&=&\rho_{1j}(0)e^{-\frac{A}{2}\tau}\cos(\Omega\tau)
-\rho_{2j}(0)e^{-\frac{A}{2}\tau}\sin(\Omega\tau),
\\  \nonumber
\rho_{2j}(\tau)&=&\rho_{1j}(0)e^{-\frac{A}{2}\tau}\sin(\Omega\tau)
+\rho_{2j}(0)e^{-\frac{A}{2}\tau}\cos(\Omega\tau),
\\
\rho_{3j}(\tau)&=&\rho_{3j}(0)e^{-A\tau}+\frac{B}{A}\rho_{0j}(0)(1-e^{-A\tau}).
\end{eqnarray}
where $A=\gamma_++\gamma_-$, $B=\gamma_+-\gamma_-$ and $\lim_{\tau\rightarrow0}\rho_{ij}(\tau)=\rho_{ij}(0)$. Eq. (\ref{solutions}) is a general analytic solution to the evolution of two-qubits system. We, therefore, can consider
the evolution of different initial states of system by choosing different state parameters $\rho_{ij}$. In this paper,
we assume that the two atoms initially share a maximally entangled state, i.e, $\rho_{00}(0)=\rho_{11}(0)=-\rho_{22}(0)=\rho_{33}(0)=1/4$,
while the rest $\rho_{ij}(0)$ vanish.

For the maximally entangled initial state given above, we find its time dependent concurrence
\begin{eqnarray}\label{concurrence}
C(\rho)=\max\big\{e^{-\frac{1}{2}A\tau}-\frac{1}{2}(1-e^{-A\tau})\sqrt{A^2-B^2},0\big\}.
\end{eqnarray}
Eq. (\ref{concurrence}) is a general expression of concurrence for the initial maximally entangled state.
For different evolutions, we can obtain different entanglements because the parameters $A$ and $B$ have
different values. In the following, we will calculate the concurrence for two special cases, one is that between a freely falling atom and its auxiliary partner in de Sitter spacetime, and the other is that between a static atom and its auxiliary partner in de Sitter spacetime.

\section{Dynamics and entanglement of freely falling atoms in de Sitter spacetime} \label{section 2}

Four-dimensional de Sitter space is most easily represented as the hyperboloid
\begin{eqnarray}\label{de Sitter hyperboloid}
z^2_0-z^2_1-z^2_2-z^2_3-z^2_4=-\alpha^2
\end{eqnarray}
embedded in five-dimensional Minkowski space with metric
\begin{eqnarray}\label{f-M-Metric}
ds^2=dz_0^2-dz_1^2-dz_2^2-dz_3^2-dz_4^2.
\end{eqnarray}
As is well known, different coordinates systems can be used to parameterize de Sitter spacetime \cite{Mottola,Birrell}. If we choose the global coordinates system $(t, \chi, \theta, \phi)$, which is
\begin{eqnarray}\label{global coordinate}
\nonumber
z_0&=&\alpha\sinh(t/\alpha),
\\ \nonumber
z_1&=&\alpha\cosh(t/\alpha)\cos\chi,
\\ \nonumber
z_2&=&\alpha\cosh(t/\alpha)\sin\chi\cos\theta,
\\ \nonumber
z_3&=&\alpha\cosh(t/\alpha)\sin\chi\sin\theta\cos\phi,
\\
z_4&=&\alpha\cosh(t/\alpha)\sin\chi\sin\theta\sin\phi,
\end{eqnarray}
then the corresponding line element is given by
\begin{eqnarray}\label{line element 1}
ds^2=dt^2-\alpha^2\cosh^2(t/\alpha)[d\chi^2+\sin\chi^2 (d\theta^2+\sin\theta^2d\phi^2)]
\end{eqnarray}
with $\alpha=\sqrt{\frac{3}{\Lambda}}$, where $\Lambda$ is the cosmological constant. If $-\infty<t<\infty, 0\leq\chi\leq\pi, 0\leq\theta\leq\pi, 0\leq\phi\leq2\pi$, it is easy to find that the coordinates cover the whole de Sitter manifold \cite{Mottola,Allen,Birrell}. The metric (\ref{line element 1}) is that of a $K=+1$ (closed) Robertson-Walker spacetime describing an expanding universe \cite{Birrell}. Under this coordinate system the freely falling atom is comoving with the expansion due to $\tau=t$ (i.e., the proper time of the freely falling observer is the same with the coordinate time), it means that the freely falling atom is moving away from the point $\chi=0$ (the center of the universe). Thus, in our model the freely falling atom moves with the trajectory (\ref{global coordinate}) from the point $(t, \chi, \theta, \phi)$ to $(t', \chi', \theta, \phi)$ in de Sitter spacetime.

We will consider the dynamic evolution of the concurrence between
two special atoms, one of which is stationary, possessed by an observer who lives at
$\chi=0$ (the atom he or she possessed is considered to be the auxiliary system discussed above), while the other is a freely falling atom interacting with
a quantized conformally coupled massless scalar field in de Sitter spacetime. To achieve this target, we must calculate the Wightman function of the massless scalar field shown above. After canonically quantizing the scalar field with the above metric (\ref{line element 1}) \cite{Allen,Allen1,Bunch,Mishima}, for the massless scalar field it is easy to find the Wightman function for the freely falling atom in the conformal coupling limit, it is
\begin{eqnarray}\label{Wightman function 1}
G^+(x-x')=-\frac{1}{16\pi^2\alpha^2\sinh^2(\frac{\tau-\tau'}{2\alpha}-i\epsilon)}.
\end{eqnarray}
Then for this Wightman function $\gamma_+$, $\gamma_-$ and $H_{eff}$ are given by
\begin{eqnarray}\label{gamma1}
\nonumber
\gamma_+&=&-\frac{\mu^2}{2\pi}\omega_0\frac{1}{1-e^{2\pi\alpha\omega_0}},
\\ \nonumber
\gamma_-&=&\frac{\mu^2}{2\pi}\omega_0\frac{1}{1-e^{-2\pi\alpha\omega_0}},
\\
H_{eff}&=&\frac{1}{2}\{\omega_0+\mu^2\mathrm{Im}(\Gamma_++\Gamma_-)\}\sigma_z
\nonumber \\
&=&\frac{1}{2}\big\{\omega_0+\frac{\mu^2}{4\pi^2}\int^\infty_0d\omega
P(\frac{\omega}{\omega+\omega_0}-\frac{\omega}{\omega-\omega_0})(1+\frac{2}{e^{2\pi\alpha\omega_0}-1})\big\}
\sigma_z,
\end{eqnarray}
where the last term of $H_{eff}$ represents the Lamb shift of the freely falling atom in de Sitter spacetime. Obviously, it is logarithmically divergent, but this divergence can be removed by introducing a cutoff on the upper limit of the integral, which is not the scope of our paper, because we have shown that the transition rates, steady states and concurrence are independent on it. It is interesting to note that the spontaneous emission of a two-level atom is $\gamma_0=\lim_{\alpha\rightarrow\infty}\gamma_-=\frac{\mu^2\omega_0}{2\pi}$, which results from the interaction of atom and external field.

\subsection{Transition rate and steady state for freely falling atom}

Let us now note that the transition rates of the freely falling atom, Eq. (\ref{gamma1}), can be rewritten as
\begin{eqnarray}\label{transition1}
\left(
  \begin{array}{c}
    \gamma_- \\
    \gamma_+ \\
  \end{array}
\right)
=\gamma_0
\left(
  \begin{array}{c}
    1+n \\
    n \\
  \end{array}
\right),
\end{eqnarray}
where $n=\frac{1}{e^{2\pi\alpha\omega_0}-1}$ is the Bose-Einstein occupation number. Nonzero $\gamma_+$ means that the freely falling atom in de Sitter spacetime, unlike the inertial atom coupled to a massless scalar field in Minkowski vacuum, has the possibility to jump from the ground state to its excited state, i.e., it has detected a quantum if it were treated as a Unruh-DeWitt detector. Further analysis shows that the ratio of the transition rates is $\frac{\gamma_+}{\gamma_-}=e^{-2\pi\alpha\omega_0}$, which obviously is the Boltzmann factor with a temperature $T_s=1/2\pi\alpha$. It signifies a thermal equilibrium distribution over the levels of the freely falling atom in de Sitter spacetime, as well as a thermal equilibrium between the freely falling atom and the external thermal field that it interacts with. Therefore, the physically acceptable de Sitter-invariant vacuum
state of the free scalar field in de Sitter spacetime is a thermal state of temperature $1/2\pi\alpha$ inside the cosmological horizon with the de Sitter boost generator fixing the horizon as the Hamiltonian, and the thermal nature of de Sitter spacetime is revealed.

As discussed above, if the transition process persists for sufficiently long time, $\tau\gg1/(\gamma_++\gamma_-)$,
the ratio of population of the atom in its ground state and excited state reaches a steady value, and then the steady state of the atom, from Eq. (\ref{time-dependent state}) and Eq. (\ref{gamma1}), is found to be a thermal state
\begin{eqnarray}\label{thermal state1}
\rho_f(\infty)=\frac{e^{-\beta_fH_s}}{Tr[e^{-\beta_fH_s}]},
\end{eqnarray}
where $\beta_f=1/T_f=2\pi\alpha$. Eq. (\ref{thermal state1}) is similar to the steady state that a two-level atom coupled to a massless scalar thermal field with temperature $1/2\pi\alpha$ in Minkowski spacetime is driven to in the infinity limit of evolution time. Therefore, a freely falling two-level atom, which is coupled to the de Sitter-invariant vacuum massless scalar field, is driven to a thermal state with Gibbons-Hawking temperature $T_f=1/2\pi\alpha$, regardless of its initial state. This thermalization phenomenon is the most obvious manifestation of the Gibbons-Hawking effect in the framework of open quantum system dynamics.

\subsection{Entanglement between freely falling atom and its auxiliary partner}

We can also understand the above thermalization process from the perspective of quantum information theory. The open quantum system, i.e, the freely falling two-level atom, is in a thermal environment, that relates to the nature of de Sitter spacetime. It is because of this thermal environment that the quantum system is subjected to a dissipation and finally is driven to an equilibrium with the dissipative medium. During the dissipation, a lot of information of the atom, e.g, quantum phase, will be lost. Therefore, one may naturally expect that the nature of de Sitter spacetime also affects the correlations of quantum systems which are placed in this spacetime.

With this purpose, We assume that the freely falling atom and its auxiliary partner initially share a maximally entangled state given by $\rho(0)=\frac{1}{4}(\sigma_0\otimes\sigma_0+\sigma_1\otimes\sigma_1-\sigma_2\otimes\sigma_2
+\sigma_3\otimes\sigma_3)$. In our model only the freely falling atom interacts with the external field, i.e., subjected to the effect of de Sitter spacetime, while its auxiliary partner is isolated from the environment. It means that only the freely falling atom is evolving as time goes on. Here it is interesting to note that such a model is similar to the free field model generally considered in relativistic quantum information, in which one usually assumes that one of the observers, called Rob, moves with a uniform acceleration, while the other observer, called Alice, stays inertial, which means that only the quantum state detected by Rob is needed to be transformed due to motion \cite{Fuentes,Tessier}, while  the quantum state detected by Alice is unchanged. Thus in our model the time-dependent state shared by the freely falling atom and its auxiliary partner, according to Eq. (\ref{solutions}), can be written as
\begin{eqnarray} \label{tdstate}
\nonumber
\rho(\tau)&=&\frac{1}{4}\bigg[\sigma_0\otimes\sigma_0
+e^{-\frac{\gamma_++\gamma_-}{2}\tau}\big(\sigma_1\otimes\sigma_1+\sigma_1\otimes\sigma_2
+\sigma_2\otimes\sigma_1-\sigma_2\otimes\sigma_2+\sigma_3\otimes\sigma_3\big)
\\
&&+\frac{\gamma_+-\gamma_-}{\gamma_++\gamma_-}\big(1-e^{-(\gamma_++\gamma_-)\tau}\big)
\sigma_3\otimes\sigma_0\bigg],
\end{eqnarray}
where the detailed formulas of $\gamma_+$ and $\gamma_-$ are shown in Eq. (\ref{gamma1}). Obviously, due to $\gamma_++\gamma_->0$, the factor $e^{-(\gamma_++\gamma_-)\tau}$in state density decreases with the increase of evolution time, which implies that there is a dissipation, and such dissipation may induce the entanglement of quantum system (the freely falling atom plus its auxiliary partner) to change. To find out how the entanglement is changing, we calculate the concurrence of the state (\ref{tdstate}), it is given by
\begin{eqnarray}\label{concurrence1}
C(\rho)_f=\max\bigg\{e^{-\frac{1}{2}\gamma_0\tau\big(\frac{e^{2\pi\alpha\omega_0}+1}
{e^{2\pi\alpha\omega_0}-1}\big)}
-\frac{1}{2}\big(1-e^{-\gamma_0\tau\big(\frac{e^{2\pi\alpha\omega_0}+1}
{e^{2\pi\alpha\omega_0}-1}\big)}\big)\cosh^{-1}(\pi\alpha\omega_0),0\bigg\}.
\end{eqnarray}

\begin{figure}[ht]
\centering
\includegraphics[scale=0.90]{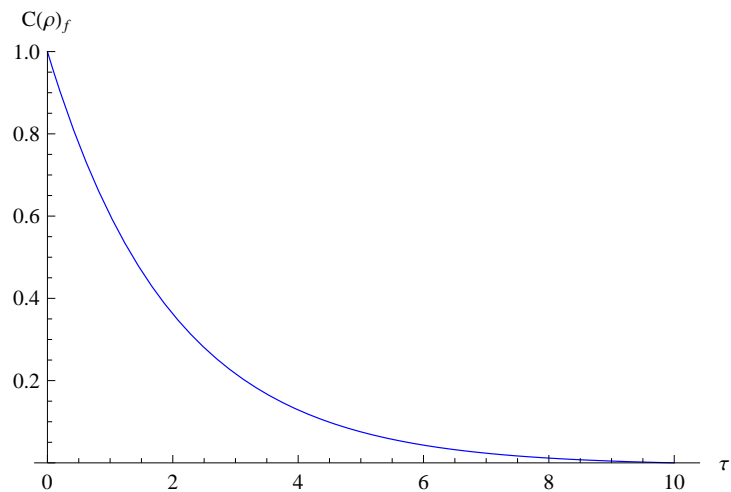}
\caption{The concurrence is plotted as a function of proper time $\tau$ (in units of $\gamma_0^{-1}$),
Gibbons-Hawking temperature $T_f=\frac{1}{2\pi\alpha}=0.1$ (in units of $\omega_0$) is assumed.}\label{f1}
\end{figure}

The concurrence (\ref{concurrence1}) is plotted in Fig. \ref{f1}. We observe that the entanglement monotonously decreases as the proper time goes on, and eventually disappears at a fixed time point
\begin{eqnarray}\label{time taken1}
\tau_0=-\frac{2}{A}\ln\big(\sqrt{\cosh^2(\pi\alpha\omega_0)+1}-\cosh(\pi\alpha\omega_0)\big).
\end{eqnarray}
This is because the freely falling atom interacts with a thermal bath, and this thermal bath makes the
freely falling atom subject to a dissipation. This dissipation will change the possibility of the  population of the upper and lower states for the freely falling atom. And due to that the freely falling atom is entangled with its auxiliary partner, the dissipation induces the entanglement to decrease. Therefore, we arrive at the conclusion that the thermal nature of de Sitter spacetime, that is felt by the freely falling atom, induces the entanglement shared by the freely falling atom and its auxiliary partner to decrease. Furthermore, in the infinity limit of evolution time, we find the asymptotic state
\begin{eqnarray}\label{asymptotic state1}
\rho(\infty)=\lim_{\tau\rightarrow\infty}\rho(\tau)=\frac{\sigma_0-\tanh(\pi\alpha\omega_0)\sigma_3}{2}
\otimes\frac{1}{2}\sigma_0.
\end{eqnarray}
Obviously, it is a separable state, more precisely, a classical state.

%%%%%%%%%%%%%%%%%%%%%%%%%%%%%%%%%%%%%%%%%%%%%%%%%%%%%%%%%%%%%%%%%%%%%%%%%%%

\section{Dynamics and entanglement of static atoms in de Sitter spacetime }\label{section 3}
Next we will discuss the dynamic evolution of the concurrence between the auxiliary atom and a static atom which interacts with a conformally coupled massless scalar field in de Sitter spacetime. For this purpose, we choose these transformations
\begin{eqnarray}\label{static coordinate}
\nonumber
z_0&=&(\alpha^2-r^2)^{1/2}\sinh(\widetilde{t}/\alpha),
\\ \nonumber
z_1&=&(\alpha^2-r^2)^{1/2}\cosh(\widetilde{t}/\alpha),
\\ \nonumber
z_2&=&r\sin\theta\cos\phi,
\\ \nonumber
z_3&=&r\sin\theta\sin\phi,
\\
z_4&=&r\cos\theta,~~0\leq r<\infty,
\end{eqnarray}
which only cover the half of the de Sitter manifold with $z_0+z_1>0$, just like the Rindler wedge. In these coordinates the line element becomes
\begin{eqnarray}\label{static coordinate1}
ds^2=\big(1-\frac{r^2}{\alpha^2}\big)d\widetilde{t}^2-\big(1-\frac{r^2}{\alpha^2}\big)^{-1}dr^2
-r^2(d\theta^2+\sin^2\theta d\phi^2).
\end{eqnarray}
Obviously, the line element (\ref{static coordinate1}) is static and possesses a coordinate singularity at $r=\alpha$, which is the event horizon for an observer situated at $r=0$. An atom at rest in this static coordinates system has the proper acceleration
\begin{eqnarray}\label{proper acceleraion}
a=\frac{r}{\alpha^2}\big(1-\frac{r^2}{\alpha^2}\big)^{-1/2}
\end{eqnarray}
to avoid falling into the horizon. Besides, the relation between the static and global coordinates system is
\begin{eqnarray}\label{static and global relation}
r=\alpha\cosh(t/\alpha)\sin\chi,~~~~\tanh(\widetilde{t}/\alpha)=\tanh(t/\alpha)\sec\chi.
\end{eqnarray}
It is of interest to note that the worldline $r=0$ in the static coordinate coincides with the worldline $\chi=0$ in the global coordinate, and an atom at rest with $r\neq0$ in the static coordinate will be accelerated relative to the observer at rest in the global coordinate with $\chi=0$. Thus, for the static atom, it has a proper time $\tau=\sqrt{g_{00}}\widetilde{t}$ and moves with the trajectory (\ref{static coordinate}) from the point $(\widetilde{t}, r, \theta, \phi)$ to $(\widetilde{t}', r, \theta, \phi)$ in de Sitter spacetime.

By solving the field equation in the static coordinates system, a set of modes will be obtained \cite{Mishima,Polarski1,Polarski2,Nakayama}. In a de Sitter-invariant vacuum, we can calculate the Wightman function for the massless conformally coupled scalar field, and which is given by \cite{Polarski3,Galtsov}
\begin{eqnarray}\label{wightman function 2}
G^+(x-x')=-\frac{1}{8\pi^2\alpha^2}\frac{\cosh(\frac{r^*}{\alpha})\cosh(\frac{{r^*}'}{\alpha})}
{\cosh(\frac{\widetilde{t}-{\widetilde{t}}'}{\alpha}-i\epsilon)
-\cosh(\frac{r^*-{r^*}'}{\alpha})},
\end{eqnarray}
with $r^*=\frac{\alpha}{2}\ln\frac{\alpha+r}{\alpha-r}$. So, for a static atom, it can be simplified to
\begin{eqnarray}\label{wightman function 3}
G^+(x-x')=-\frac{1}{16\pi^2\kappa^2\sinh^2(\frac{\tau-\tau'}{2\kappa}-i\epsilon)},
\end{eqnarray}
where $\kappa=\sqrt{g_{00}}\alpha$. Comparing the Wightman function (\ref{wightman function 3}) with (\ref{Wightman function 1}), it is easy to obtain
\begin{eqnarray}\label{gamma2}
\nonumber
\gamma_+&=&-\frac{\mu^2}{2\pi}\omega_0\frac{1}{1-e^{2\pi\kappa\omega_0}},
\\ \nonumber
\gamma_-&=&\frac{\mu^2}{2\pi}\omega_0\frac{1}{1-e^{-2\pi\kappa\omega_0}},
\\
H_{eff}&=&\frac{1}{2}\{\omega_0+\mu^2\mathrm{Im}(\Gamma_++\Gamma_-)\}\sigma_z
\nonumber \\
&=&\frac{1}{2}\big\{\omega_0+\frac{\mu^2}{4\pi^2}\int^\infty_0d\omega
P(\frac{\omega}{\omega+\omega_0}-\frac{\omega}{\omega-\omega_0})(1+\frac{2}{e^{2\pi\kappa\omega_0}-1})\big\}
\sigma_z.
\end{eqnarray}
for the atom at rest in the static coordinate system. Again, $\lim_{\kappa\rightarrow\infty}\gamma_-=\frac{\mu^2}{2\pi}\omega_0$ is the spontaneous emission due to the quantum
interactions of a two-level atom with the massless scalar field.

\subsection{Transition rate and steady state for static atom}

As shown in Eq. (\ref{gamma2}), the transition rates of the static atom, which is in interaction with the
conformally coupled massless scalar field in de Sitter-invariant vacuum, can be rewritten as
\begin{eqnarray}\label{transition2}
\left(
  \begin{array}{c}
    \gamma_- \\
    \gamma_+ \\
  \end{array}
\right)
=\gamma_0
\left(
  \begin{array}{c}
    1+n \\
    n \\
  \end{array}
\right),
\end{eqnarray}
where $n=\frac{1}{e^{2\pi\kappa\omega_0}-1}$ represents the Bose-Einstein occupation number. Nonzero $\gamma_+$ represents that
the static atom in de Sitter spacetime, unlike the inertial atom coupled to a massless scalar field in Minkowski vacuum, can
absorb energy from the environment, and which induces the static atom to jump from the ground state to its excited state. $\gamma_-$ indicates the emission of the static atom. Due to that the excitation of the two-level detector can occur only if there are particles in the field to which it is coupled, nonzero $\gamma_+$ here also signals the presence of particles as seen by the static observer in de Sitter spacetime, and the departure from the de Sitter-invariant vacuum state. Further analysis shows that the ratio of the transition rates is $\frac{\gamma_+}{\gamma_-}=e^{-2\pi\kappa\omega_0}$, which is the Boltzmann factor and is identical to the detailed balance relation for transition rates in a thermal environment at temperature $T_s=1/2\pi\kappa$. Therefore, we arrive at the conclusion that the static atom in de Sitter spacetime feels thermal bath of particles with temperature $T_s=1/2\pi\kappa$, and the temperature should be considered as corresponding to the actual physical temperature of the environment as seen by the observer.

Similar to the analysis of the freely falling atom, if $\tau\gg1/(\gamma_++\gamma_-)$, i.e., the transition process persists for sufficiently long time, the ratio of population of the static atom in its ground state and excited state will also reach a steady value. From Eq. (\ref{time-dependent state}) and Eq. (\ref{gamma1}), we find that its steady state is a thermal state
\begin{eqnarray}\label{thermal state2}
\rho_s(\infty)=\frac{e^{-\beta_sH_s}}{Tr[e^{-\beta_sH_s}]},
\end{eqnarray}
where $\beta_s=1/T_s=2\pi\kappa$. It is interesting to note that $T^2_s=T^2_f+T^2_U$ with the Unruh temperature $T_U=a/2\pi$.
Eq. (\ref{thermal state2}) is similar to the steady state that a two-level atom coupled to a massless scalar thermal field with temperature $1/2\pi\kappa$ in Minkowski spacetime will be driven to after a sufficiently long time of evolution. Therefore, a static two-level atom, which is coupled to the de Sitter-invariant vacuum massless scalar field, is driven to a thermal state with temperature $T_s=1/2\pi\kappa$, regardless of its initial state. This thermalization phenomenon, from the perspective of a static observer, is the most obvious manifestation of the thermal nature of de Sitter spacetime in the framework of open quantum system dynamics.

\subsection{Entanglement between static atom and its auxiliary partner}

To study how the thermal nature of de Sitter spacetime, which is considered as a thermal environment from the view of the static atom, affects the entanglement shared by the static atom and its auxiliary partner,
we also assume that the freely falling atom and its auxiliary partner initially share a maximally entangled state given by $\rho(0)=\frac{1}{4}(\sigma_0\otimes\sigma_0+\sigma_1\otimes\sigma_1-\sigma_2\otimes\sigma_2
+\sigma_3\otimes\sigma_3)$. Analogously, only the static atom interacts with the external field, i.e., subjected to the effect of de Sitter spacetime it feels, while its auxiliary partner is isolated from the environment. For this case, according to (\ref{solutions}), (\ref{concurrence}) and (\ref{gamma2}) the concurrence of the evolving quantum state of the static atom and its auxiliary partner is
\begin{eqnarray}\label{concurrence2}
C(\rho)_s=\max\bigg\{e^{-\frac{1}{2}\gamma_0\tau\big(\frac{e^{2\pi\kappa\omega_0}+1}
{e^{2\pi\kappa\omega_0}-1}\big)}
-\frac{1}{2}\big(1-e^{-\gamma_0\tau\big(\frac{e^{2\pi\kappa\omega_0}+1}
{e^{2\pi\kappa\omega_0}-1}\big)}\big)\cosh^{-1}(\pi\kappa\omega_0),0\bigg\}.
\end{eqnarray}

\begin{figure}[ht]
\centering
\includegraphics[scale=1.0]{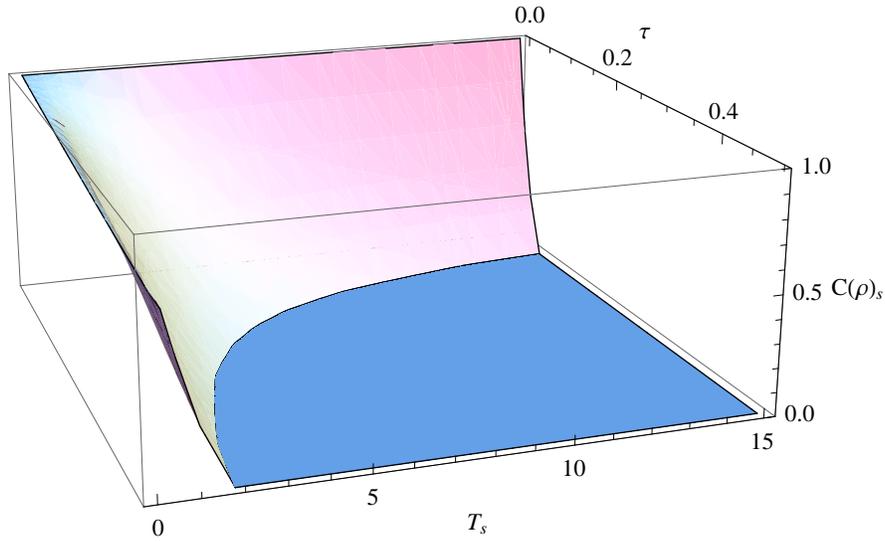}
\caption{The concurrence is plotted as a function of proper time $\tau$ (in units of $\gamma_0^{-1}$) and thermal
temperature $T_s=\frac{1}{2\pi\kappa}$ (in units of $\omega_0$).}\label{f2}
\end{figure}

The concurrence (\ref{concurrence2}) is plotted in Fig. \ref{f2} as a function of proper time $\tau$ of the static atom and the thermal temperature $T_s$ felt by it. We can see from Fig. \ref{f2} that the initial entanglement decreases monotonously with the
the increase of the proper time and the thermal temperature $T_s$. The greater the thermal temperature is, the earlier
the initial entanglement disappears, and at a fixed thermal temperature the time taken, $\tau_0$, for the system to completely disentangle is given by
\begin{eqnarray}\label{time taken2}
\tau_0=-\frac{2}{A}\ln\big(\sqrt{\cosh^2(\pi\kappa\omega_0)+1}-\cosh(\pi\kappa\omega_0)\big).
\end{eqnarray}
This is because the static atom in de Sitter spacetime feels as if it is in a thermal bath, and due to the interaction with the thermal bath which plays a dissipative role during the evolution of the static atom, the possibility of the  population of the upper and lower states of the static atom will be changed. As a result of that, the entanglement of the whole quantum system, the static atom plus its auxiliary partner, is subjected to a decrease. Thus, we arrive at the conclusion that the thermal nature of de Sitter spacetime, that is felt by the static atom, induces the entanglement shared by the static atom and its auxiliary partner to decrease and finally disappears. Furthermore, for the dimensionless parameter $\xi=\kappa\omega_0\ll1$,
i.e, $\omega_0\ll\frac{1}{\kappa}$, which corresponding to the case that the static atom locates near the horizon because of
$\kappa=\sqrt{\alpha^2-r^2}$, then we find
\begin{eqnarray}\label{time limit}
\tau_0=\ln(\frac{1}{\sqrt{2}-1})\frac{1}{T_s}+O(T^{-3}_s).
\end{eqnarray}
Thus when the static atom stays near the universe horizon, the proper time taken to disentangle it and its auxiliary partner is proportional to the inverse of the thermal temperature $T_s$. Here, it is needed to note that in this case, $T_U\gg T_f$, so we have $T_s\approx T_U$, and the dissipative effect can be thought to completely come from the Unruh effect.

%%%%%%%%%%%%%%%%%%%%%%%%%%%%%%%%%%%%%%%%%%%%%%%%%%%%%%%%%%%%%%%%%%%%%%%%%%%%%%%%%%%%%%%%%%%%%%%%%%%%%%%%%%

\section{Discussions}\label{section 4}

No matter for the freely falling or the static case, a single atom can be understood as a single detector. It is clearly shown that in terms of the transition rate and steady state both the freely falling and the static two-level atoms, which interact with a conformally coupled massless scalar field in the de Sitter spacetime, are in structural similarity to that of an inertial atom immersed in a thermal bath in the Minkowski spacetime. Thus, from the transition rate, steady state and so on of a single atom, it is not possible to distinguish between de Sitter spacetime and a thermal bath. However, as shown is Ref. \cite{Steeg}, using two inertial detectors, interactions with the field in the thermal case will entangle certain detector pairs that would not become entangled in the corresponding de Sitter case, thus one can tell the difference between the thermal Minkowski spacetime and de Sitter universe by the entangling power of two detectors. Unlike that in Ref. \cite{Steeg} where both the detectors interact with the quantum fields and are comoving with the expansion of universe, in our model we assume that only the freely falling or the static atom interacts with the external field, while its auxiliary partner is isolated. It means that only one party of the bipartite quantum system (the freely falling or the static atom plus its auxiliary partner) evolves with time. In this regard, let us note that this model, in some sense, also can be understood as a single detector case. Thus, the results we obtained are the same with that of the thermal bath case. However, this model is similar to the free field model generally considered in relativistic quantum information, where one usually assumes that one of the observers, called Rob, moves with a uniform acceleration, while the other observer, called Alice, stays inertial \cite{Fuentes,Tessier}. Thus, our model allows us to consider the evolution of entanglement shared by two relatively moved observers in de Sitter spacetime, and discuss how the nature of de Sitter spacetime affects quantum teleportation such as Ref. \cite{Feng}.

Because the dynamics of both the freely falling and static atoms in de Sitter spacetime are in structural similarity to that of an inertial atom immersed in a thermal bath in the Minkowski spacetime, it is possible to allow us to simulate the dynamics of such quantum systems in the table-top experiment. Based on the well established experimental techniques \cite{Leibfried}, and repeating the same analysis in Ref. \cite{Marco}, the freely falling atom and the static atom in de Sitter spacetime studied above can be simulated by trapped ions or circuit QED. And during this process a little of differences are that we have to replace the Unruh temperature by the temperatures $T_f=1/2\pi\alpha$ for the freely falling atom and $T_f=1/2\pi\kappa$ for the static atom. After doing like that, the transition probability and the steady state of the simulation quantum system may allow us to understand the dynamics of the freely falling and static atoms in de Sitter spacetime from the view of table-top simulation experiment.

As discussed above, the thermal baths felt by the freely falling atom (the static atom) will drive
the entanglement between it and its auxiliary partner to sudden death, this is because the thermal baths can be thought of as external environments, which will induce quantum decoherence in the quantum system (freely falling atom + auxiliary atom or static atom + auxiliary atom). Similar to the model constructed in \cite{Ting}, while assuming the field is thermal and only one of the subsystems interacts with the external field, one can construct a simulation quantum system to study the decoherence of the freely falling atom (static atom) and its auxiliary partner in de Sitter spacetime, and the disentanglement can be detected by the means proposed in \cite{Santos}. Thus the decrease and sudden death of entanglement, in principle, could provide us an indicator to estimate whether the quantum system feels thermal or non-thermal in the simulated experiment, as well as in de Sitter spacetime.

%%%%%%%%%%%%%%%%%%%%%%%%%%%%%%%%%%%%%%%%%%%%%%%%%%%%%%%%%%%%%%%%%%%%%%%%%%%%%%%%%%%%%%%%%%%%%%%%%%%%%%%%%%

\section{Conclusions}\label{section 5}

In the framework of open quantum systems, we find that the dynamics of both the freely falling and static two-level atoms, which interact with a conformally coupled massless scalar field in the de Sitter spacetime, is in structural similarity to that of an inertial atom immersed in a thermal bath in the Minkowski spacetime, which reveals the thermal nature of de Sitter spacetime from a different physical context. We simultaneously show that the thermal baths, that are felt by the freely falling and static atoms, can be thought of as the thermal environments for the quantum systems, the freely falling atom (the static atom) plus its auxiliary partner. However, in these systems, only the freely falling atom (the static atom) interacts with the thermal noise, while its auxiliary partner is isolated from it. Then the nature of de Sitter spacetime can affect the dynamic evolution of entanglement for two-level atoms by affecting the dynamic evolution of the freely falling atom (the static atom).

For the freely falling atom, the entanglement between it and its auxiliary partner decreases as time goes on, and eventually
vanishes at a fixed time. This is because the freely falling atom feels a Gibbons-Hawking thermal bath, and this thermal bath, through affecting the freely falling atom, induces the entanglement to decay.
For the static atom, it also feels a thermal bath that results from both the Gibbons-Hawking effect and the Unruh effect associated with the static atom's proper acceleration, and this thermal bath equally causes the entanglement between the static atom and its auxiliary partner to degenerate. Furthermore, the greater the temperature that the static atom feels, the earlier the entanglement disappears. When the static atom stays near the universe horizon, the proper time taken to disentangle it and its auxiliary partner is proportional to the inverse of the temperature it feels.

\begin{acknowledgments}

This work was supported by the  National Natural Science Foundation
of China under Grant Nos. 11175065, and 10935013; the National Basic
Research of China under Grant No. 2010CB833004; the SRFDP under
Grant No. 20114306110003; PCSIRT, No. IRT0964; Hunan Provincial Innovation Foundation For Postgraduate under Grant No CX2012B202; the Hunan Provincial Natural Science Foundation of China under Grant No 11JJ7001;  and
Construct Program of the National Key Discipline.

\end{acknowledgments}

%\newpage

\end{document}